\documentstyle[preprint,prl,aps,epsfig]{revtex}
\begin{document}

\title{Does particle decay cause wave function collapse:\break
An experimental test}

\author{Spencer R. Klein$^1$ and Joakim Nystrand$^2$} 
\address{$^1$ Lawrence Berkeley National Laboratory, Berkeley, CA 94720 
\newline $^2$ Department of Physics, Lund University, Lund, Sweden} 

\break 
\maketitle

\begin{abstract}

We describe an experimental test of whether particle decay causes wave
function collapse.  The test uses interference between two well
separated, but coherent, sources of vector mesons.  The short-lived
mesons decay before their wave functions can overlap, so any
interference must involve identical final states.  Unlike previous
tests of nonlocality, the interference involves continuous variables,
momentum and position.  Interference can only occur if the wave
function retains amplitudes for all possible decays.  The interference
can be studied through the transverse momentum spectrum of the
reconstructed mesons. 

\end{abstract}

\narrowtext

In 1935, Einstein, Podolsky and Rosen (EPR) showed that quantum
mechanics required that wave functions can be non-local\cite{EPR}.
When a system is observed, the wave function collapses from one which
contains amplitudes for a host of possible outcomes to smaller set of
possibilities, in accord with the measurement.  This collapse is
instantaneous; much has been written about its superluminous
nature. Most studies of the EPR paradox have tested Bell's
inequality\cite{bell} using spin correlations, usually with photons
produced in pairs\cite{pra}.  Experimenters measure the spin
correlations using two polarizers with a varying angle between them.
Bell found that models with non-local wave functions and models with
hidden variables produced different angular correlation spectra.
Previous tests of non-locality used discrete variables like
'pseudo-spin' for CP violation, as with studies using the reaction
$\Phi\rightarrow K^+K^-$\cite{cplear}.

We describe a very different system that, in contrast to the $K_LK_S$
system, is sensitive to the collapse of continuous variables in a wave
function\cite{parkins}.  Short-lived vector mesons (VMs) are produced
with a fixed phase relationship at two separated sources.  Even though
the mesons do not come from a single source, and, in fact, share no
common history\cite{science}, the system acts as an interferometer.
The meson lifetimes are short compared to the source separation, so
the mesons decay before their wave functions can spatially overlap.

Any interference between the two sources must involve the decay
products.  Interference is only possible between identical final
states.  With the large phase space for final states, interference can
only occur if the wave functions retain amplitudes for all possible
decay channels and angular distributions long after the decay takes
place\cite{prb}.  We have previously calculated the interference
pattern\cite{usint}.  This letter will focus on the effects of the
wave function collapse and Bells inequality-like tests, and sketch an
alternate derivation of the interference, to emphasize the symmetries
of the system.

Figure 1 shows electromagnetic VM production in relativistic heavy ion
collisions at large impact parameters, $\vec{b}$.  A photon from the
electromagnetic field of one nucleus fluctuates to a virtual
quark-anti-quark pair which elastically scatters from the other
nucleus, emerging as a real vector meson\cite{review}.  Either nucleus
can emit the vector meson.  The momentum transfers from the nuclei are
similar, and they remain in the ground state, so it is impossible to
determine which nucleus emitted the photon and which is the target.

The electromagnetic interaction (photon) has a long range, while the
elastic scattering has a short range, around 0.6 fm\cite{muller}, far
smaller than the $< 7$ fm radius of a heavy nucleus or the
typical impact parameter.  So, VM production takes place
essentially `on top of' the emitting nucleus, and the two nuclei act
as a two-source interferometer.

Electromagnetic VM production is studied at the Relativistic Heavy Ion
Collider (RHIC) at Brookhaven National Laboratory, where gold ions
collide at center of mass energies up to 200 GeV per nucleon.
Starting in 2006, the Large Hadron Collider (LHC) at CERN will collide
lead ions at a center of mass energy of 5.5 TeV per nucleon.

The cross sections were previously calculated\cite{usPRC} using the
Glauber approach\cite{review} with the photon spectrum given by the
Weizs{\"a}cker-Williams virtual photon method\cite{jackson}.  The
calculated photonuclear cross sections agree with data to within 20\%.
The cross sections are large, about 10\% of the hadronic cross section
at RHIC, rising to 50\% at the LHC.  The corresponding production
rates, more than 100 $\rho^0$/sec at RHIC, rising to 230,000
$\rho^0$/sec at the LHC, are large enough that it will be easy to
collect adequate statistics to study wave function collapse.  Already,
the STAR collaboration\cite{falk} has observed more than 10,000
$\rho^0$, which should be enough to observe the interference.

The impact parameters for these interaction are large compared with
the nuclear radii, $R_A$.  For $\rho$ and $\omega$ production, the
median impact parameter $\langle b \rangle$ is about 40 fm at RHIC,
rising to 300 fm at the LHC; for the $J/\psi$, $\langle b \rangle$
rises from 23 fm at RHIC to about 50 fm at the LHC. All are much
larger than $R_A\approx 7$ fm for heavy ions.  It is possible to
select events with smaller $\langle b\rangle$, but still with
$b>2R_A$, by choosing events where VM production is accompanied by
nuclear breakup\cite{breakupprl}.

The $\langle b \rangle$ are larger than the distance travelled by most
VMs before they decay.  The VM lifetimes $\tau$ range from
$4\times10^{-24}$s for the $\rho$ up to $7.5\times10^{-21}$s for the
$J/\psi$.  The VM are produced with typical transverse momentum
$p_T\approx 2\hbar/R_A \approx 60$MeV/c; at mid-rapidity, the
longitudinal momentum is zero, so VM have a median decay distance
$d=2\hbar c\tau/R_A M_V$. Except for the $J/\psi$, $d\ll \langle
b\rangle$; for the $J/\psi$ at the LHC, $d\approx \langle b\rangle$.

The final state wave function from ion source $i$ at a time $t$ can be
expressed schematically
\begin{equation}
\psi(t)_i = \exp{(-t/2\tau)} \, |V> + (1-\exp{(-t/2\tau)}) \, |DP>
\label{eq:decay}
\end{equation}
where $\tau$ is the vector meson lifetime, $|V>$ is the vector
meson wave function, and $|DP>$ is the final state.  For stable
particles, $\tau=\infty$, the decay products drop out, leaving a
conventional two-source interferometer.

The interference can be seen by examining the symmetries of the
system. The total amplitude $A_T$ for observing the VM with momentum
$\vec{p}$ at position $\vec{r}$, and time, $t$, depends on the
production amplitude $A(\vec{p},\vec{x},t')$ and a propagator
$P(\vec{p},\vec{x},t',\vec{r},t)$ which transports the meson from
$\vec{x}',t'$ to $\vec{r},t$:
\begin{equation}
A_T(\vec{p},\vec{r},t) = \int A(\vec{p},\vec{x}',t')
P(\vec{p},\vec{x},t',\vec{r},t)
d\vec{x} dt'.
\end{equation}
The production amplitude $A(\vec{p},\vec{r},t)$ depends on the
electromagnetic field, $E(\vec{x},t')$, nuclear density
$\rho(\vec{x},t)$ and the amplitude $f(\vec{p},\vec{k})$ for a photon
with momentum $\vec{k}$ 
to fluctuate to a $q\overline q$ pair and
scatter from a nucleon, emerging as a vector meson with momentum
$\vec{p}$:
\begin{equation}
A(\vec{p},\vec{x}',t') =
f(\vec{p},\vec{k}) 
\rho(\vec{x}',t') E(\vec{x}',t')
\end{equation}
The electromagnetic field at a distance $b$ from a nucleus is a
Lorentz-contracted pulse with a width $b/\gamma$ where $\gamma$ is the
Lorentz boost. When $\gamma\gg 1$, the electric and magnetic
fields are perpendicular and the overall field may be represented as a
stream of almost-real photons, with energies up to
$\hbar\gamma/b$\cite{jackson}.  The photon amplitude is proportional
to $E(\vec{x}',t')$.  The scattering amplitude is obtained
from data; only its symmetries are important here.  Absorption of the
nascent $\rho^0$ is neglected, but could be included with an
additional position-dependent variable, effectively modifying
$\rho(x,t')$.

At large distances, the propagator for a VM with energy
$\omega = \sqrt{M_V^2 + |\vec{p}|^2}$ may be modelled with a plane
wave. Neglecting, for now, VM decays,
\begin{equation}
P(\vec{p},\vec{x},t',\vec{r},t) =
e^{i(\vec{p}\cdot(\vec{r}-\vec{x}) - \omega (t-t'))}
\end{equation}

The nuclear density is symmetric around the center of mass (origin),
giving it positive parity, while the antisymmetric electric field has
negative parity): $\rho(\vec{x},t') = \rho(-\vec{x},t')$ and
$E(\vec{x},t')= -E(-\vec{x},t')$.  With this, the range of integration
in Eq. (2) can be restricted to a single nucleus:
\begin{equation} 
A_T(\vec{p},\vec{r},t) =  \int_{y>0} 
d\vec{x} dt' 
\rho(\vec{x},t')  E(\vec{x},t')
e^{i(\vec{p}\cdot\vec{r} - \omega (t-t'))}
\big[f(\vec{p},\vec{k})e^{i\vec{p}\cdot\vec{x}} -
f(\vec{p},-\vec{k})e^{-i\vec{p}\cdot\vec{x}} \big]
\end{equation}
The only differences between the two nuclei are the phases $\pm
i\vec{p}\cdot\vec{x}$ and between $f(\vec{p},-\vec{k})$ and
$f(\vec{p},\vec{k})$.  The latter is because the sign of $p_z$ reduces
the symmetry of the system.  Of course, interference is only
significant when $|f(\vec{p},\vec{k})|\approx |f(\vec{p},-\vec{k})|$
which occurs near $p_z = 0$.

The equation simplifies by defining $\vec{x} = \vec{b}/2 + \vec{x'}$.
The bulk of the cross section is from when the photon couples
coherently to the target nucleus, {\it i.e.} when
$\vec{k}\cdot\vec{x'} \ll \hbar$, so the exponential phase is constant
over the nucleus. Then, the maximum transverse and longitudinal
momenta are $\hbar/R_A$ and $\gamma\hbar/R_A$.  Emitted photons are
subject to similar limits\cite{vidovic}.  Near $p_z=0$, the photon
momentum, and the momentum exchange due to the scattering are very
similar so it isn't possible to determine which nucleus emitted the
photon, and which was the scatterer; in fact, at $\vec{p}=0$, the two
momentum transfers are equal and opposite.

The electromagnetic pulse lasts a time, $b/c\gamma$, which may be
slightly longer than one photon period ($\hbar/\omega)$. This partial
temporal incoherence will reduce the overall production amplitude.
There is a pairwise cancellation between space-time volume elements
$d\vec{x} dt'$ at positions $\vec{x}$ and $-\vec{x}$, so the
interference is not affected.

With this, $\vec{p}\cdot\vec{x} = \vec{p}\cdot \vec{b}/2$ and
\begin{equation} 
A_T(\vec{p},\vec{r},t) =  \int_{y>0} 
d\vec{x'} dt' 
\rho(\vec{x'},t')  E(\vec{x'},t')
e^{i(\vec{p}\cdot\vec{r} - \omega (t-t'))}
\big[f(\vec{p},\vec{k})e^{i\vec{p}\cdot\vec{b}/2} -
f(\vec{p},-\vec{k})e^{-i\vec{p}\cdot\vec{b}/2} \big].
\label{eq:at}
\end{equation}
We now introduce a few approximations.  The amplitude for production
from the first nucleus is $A_1 (\vec{p},\vec{r},t) = \int_{y>0}
d\vec{x} dt' \rho(\vec{x},t') E(\vec{x},t') f(\vec{p},\vec{k})$ and
we define $c$ to be the ratio of the amplitudes for production
from the two nuclei:
\begin{equation}
c(p_z)  =
{\int_{y>0} d\vec{x} dt' \rho(\vec{x},t')
E(\vec{x},t') f(\vec{p},-\vec{k}) 
\over 
\int_{y>0} d\vec{x} dt' \rho(\vec{x},t')
E(\vec{x},t')     f(\vec{p},\vec{k})}.
\end{equation}
The ratio $f(\vec{p},-\vec{k})/ f(\vec{p},\vec{k})$ does not vary
significantly over the nucleus, so the single nucleus production
amplitude factors out of the integral in Eq. (\ref{eq:at}).  The
transverse momenta do not affect $c$, and $k_z=M_V^2/4p_z$.  Then,
\begin{equation}
A_T(\vec{p},\vec{r},t) = A_1 (\vec{p},\vec{r},t)
\bigg[e^{i\vec{p}\cdot \vec{b}/2} -c(p_z) 
e^{-i\vec{p}\cdot \vec{b}/2}\bigg].
\end{equation}
The amplitude factorizes into a magnitude and an interference term.
The $p_T$ dependence of $ A_1 (\vec{p},\vec{r},t)$ is dominated by the
nuclear form factors, with the bulk of the production having $p_T <
2\hbar/R_A$.  Most of the uncertainties discussed earlier do not
affect the interference term. The time and $z$ variation in
$E(\vec{x},t')$ should be largely independent of $k$. In the soft
Pomeron model, the photon to VM coupling increases slowly
with $k$ and has an almost constant phase.  At RHIC, a photon-meson
term is also present, but the phase of $c$ still changes only slowly
with $k$\cite{review}.

The interference is clearest when $p_z$=0. Then $c=1$ and the
approximations introduced in defining $c$ disappear.  The amplitude is
$A_T(\vec{p}, \vec{r},t) = 2iA_0 \sin{(\vec{p}\cdot \vec{b}/2)}$.  and
the cross section is
\begin{equation}
\sigma \sim |A_T(\vec{p}, \vec{r},t)|^2 = 2 A_0^2 \big[
1-\cos{(\vec{p}\cdot \vec{b})} \big].
\label{eq:sigma}
\end{equation}
This formula applies for stable particle emission, such as
bremsstrahlung photon emission in $e^-e^-$\cite{eecoll}
and $pp$ collisions\cite{nature}.

For short-lived particles, the situation is more interesting. 
We can express the cross section in Eq.~\ref{eq:sigma} as a sum of a 
coherent (interfering) and an incoherent (non-interfering) term  
\begin{equation}
2 A_0^2 \big[ 1- (1-\eta) \cdot \cos{(\vec{p}\cdot \vec{b})} \big], 
\end{equation}
where $\eta$ measures the degree of decoherence, as has been done for 
the system $\Phi\rightarrow K^0\overline{K}^0$\cite{cplear,decoherence}. 
Complete quantum mechanical coherence between the two sources here means 
$\eta = 0$. 

If the interference is restricted to the time the system spends in the 
vector meson (parent) state, then one would expect partial decoherence. 
According to Eq.~\ref{eq:decay}, only a fraction 
$\exp( - (M_V b)/(\omega \tau) )$ of the vector mesons will have 
survived long enough for the amplitude to propagate the distance b between 
the nuclei before the decay. This scenario thus corresponds to 
\begin{equation}
\eta = 1 - \exp( - \frac{M_V b}{\omega \tau} ).  
\end{equation}
For the $\rho$, $\omega$, and $\phi$, which all have $c \tau \ll <b>$, 
$\eta \approx 0$ and one expects almost complete decoherence in this scenario. 
For the $J/\Psi$, on the other hand, $c \tau \approx <b>$, so the decoherence 
would be only partial. This scenario could therefore be distinguished by 
observing the $J/\Psi$, while for the lighter vector mesons it would essentially 
be indistinguishable from a scenario with no interference. 

However, there is a broader issue. A distant observer sees the decay 
products from the original meson, and most VM have many decay modes.
The final state may be written $|DP>=\Sigma_j\alpha_j |DP_j>$, where 
$\alpha_j$ is the amplitude for final state $|DP_j>$.
For example, the $J/\psi$ can decay to $e^+e^-$ or $\pi^+\pi^-\pi^0$,
among many possibilities. The probability for any specific final state
is small, less than 7\%. 

If the wave function contains amplitudes for every possible decay mode, 
at times $t\gg \tau$, the individual particles may be modelled as plane 
waves and the wave function is 
\begin{equation}
\Psi(t) = \int_0^t  {dt_d \over 2\tau} 
\ \ \bigg[e^{i\vec{p}\cdot (\vec{x}_d + \vec{b}/2)}
+ c e^{i\vec{p}\cdot (\vec{x}_d - \vec{b}/2)} \bigg]
\ \ e^{-t_d/2\tau-i\omega t_d} 
\ \ |\Sigma_j\sqrt{Br_j} \Psi_j> 
\label{eq:wf1}
\end{equation}
where the decays occur at time $t_d$ and displacement
$\vec{x}_d=(\vec{p}/M_V)t_d$ from the production points. Here, $Br_j$ are 
the branching ratios to different final states, and $\Psi_j$ is the
$k-$particle final state
\begin{equation}
\Psi_j = \Sigma_k e^{i[\vec{p}_k (\vec{x}_k-\vec{x_d}) 
- \omega_j(t-t_d)]} |\Psi_{jk}>   
\ \delta(\Sigma_k \omega_k - \omega) 
\delta(\Sigma_k {\vec p}_k -\vec{p}).
\label{eq:wf2}
\end{equation}
Here $\vec{x}_k$, $\omega_k$ and $\vec{p}_k$ are the particle
positions, energy, and momenta, and $|\Psi_{jk}>$ includes the
particle types and angular distribution.  The $\delta$ functions
impose 4-momentum conservation.

If the decay occurs before the two amplitudes overlap, then either 
the wave function must retain amplitudes for all final states, or the 
two decays will produce different final states, and the interference 
must be small. Also, as Eq. (9) shows, amplitudes for different decay 
times must also be included. Thus, the presence or absence of 
interference tests whether particle decay collapses the wave function.

The two source terms in Eq. (\ref{eq:wf1}) entangle the final state
wave functions.  The phases differ by 
$\exp{[i(\vec{p}\cdot\vec{b}+\delta)]}$ where $\delta$ is the phase of
$c$.  Any measurement on one decay product at least partially collapse
the wave function of the others\cite{pra}.  Detectors could accurately
measure either the position or momentum of the $k$ final state
particles. 'Accurately', is compared to the relevant distance ($b$) or
momentum ($\hbar/b$) scales.  By these metrics, current and planned
experiments measure momentum accurately, but not position.

The interference pattern, Eq. (\ref{eq:sigma}) can be seen in the
reconstructed VM $p_T$\cite{usint}.  Because $b$ is not generally
measurable, $\sigma$ must be integrated over all $b$.  Figure 2 shows
the expected $p_T$ spectrum for $\rho^0$ production at mid-rapidity at
RHIC\cite{usint}.  The large dip for $p_T < \hbar/\langle b \rangle$
is a distinctive experimental signature.

Alternately, at least in a gedanken experiment, position sensitive
detectors could be used to localize the decay to a single nucleus,
provided the ion trajectories are known.  The decay $J/\psi\rightarrow
e^+e^-$ produces two relatistic electrons that are back-to-back in the
transverse plane.  For $p_T=0$, a line between the two measured
electron positions will intersect one of the ion trajectories.  When
the electron $\vec{p}_k$ are perpendicular to $\vec{b}$ then it is
possible to determine which nucleus emitted the VM.  The nonzero meson
$p_T$ introduces some uncertainty, but not enough to encompass both
ion trajectories.  For detectors 500 fm from the collision point, the
pointing uncertainty is 16 fm, less than the $\langle b\rangle\sim 50$
fm for $J/\psi$ at the LHC.

As with existing tests of Bell's inequality, two detectors, on
opposite sides of the collision region could randomly measure either
position or momentum. All single-detector measurements are insensitive
to what happens in the other detector.  However, when both detectors
measured position, the production point could be determined, while
when both measured momentum, a null at $p_T=0$ would be seen, showing
the interference.  These two possibilities are only compatible if the
wave function collapses when a measurement is made, and not earlier,
when the meson decays.  For $J/\psi\rightarrow e^+e^-$, the collapse
would have to be superluminal.

One possible source of decoherence is the decay timing. The VM are 
produced nearly at rest, but the decay products may be relativistic.  
The maximum flight time difference from the two sources to a detector 
is $b/c$.  If the detectors could resolve this time difference, this 
could partially localize the production, reducing the coherence.

Since the probability of producing a VM in grazing collisions
($b=2R_A$) is high, about 1\% at RHIC and 3\% at the LHC\cite{usPRC},
multiple VM production is also observable.  Multi-meson final states
will exhibit more complicated entanglements, with possibly new
behavior.

We have described a 2-source interferometer for short lived particles,
and showed that its description requires a non-local wave function.
The observation of interference will clearly demonstrate that, after a
decay, a systems' wave function includes amplitudes for all possible
decay modes and angular distributions, and does not collapse to a
specific decay mode until the wave function is externally observed.
Measurements of this interference should be available soon.  The STAR
detector has observed exclusive $\rho^0$ production in gold collisions
at RHIC\cite{falk}.  Current data should provide higher statistics and
an accurate $\rho^0$ $p_T$ spectrum, probing the EPR paradox for
continuous variables.

This work was supported by the U.S. Department of Energy under
Contract No. DE-AC-03076SF00098 and by the Swedish Research
Council (VR).

\begin{figure}
\setlength{\epsfxsize=0.75\textwidth}
\centerline{\epsffile{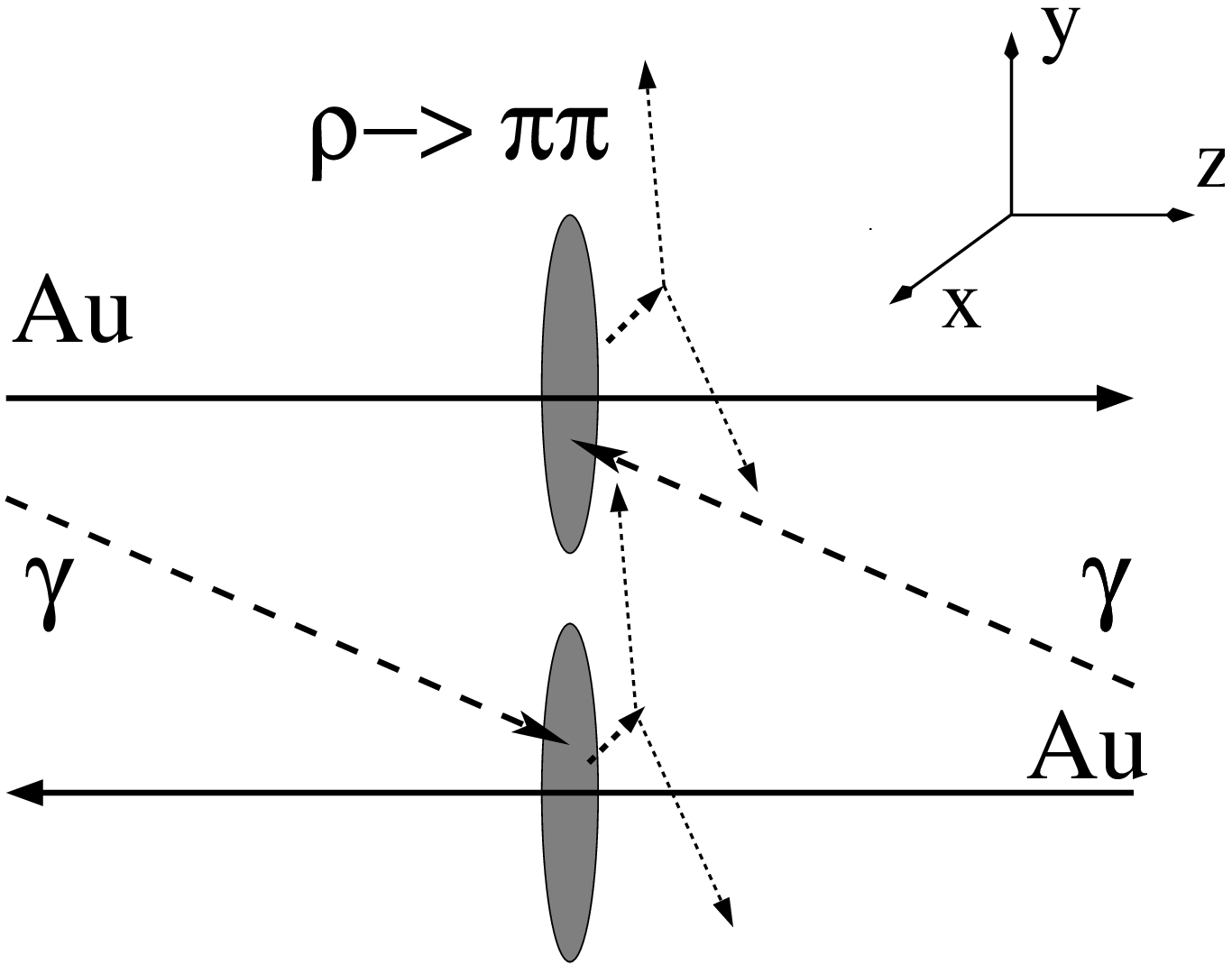}}
\caption{Diagram showing ultra-peripheral $\rho^0$ production and
decay in heavy ion collisions.  The nuclear momenta follow the $z$
axis, and come closest at $z=0$, when their separation (impact
parameter), $\vec{b}$ follows the $y$ axis.}
\label{diag}
\end{figure}

\begin{figure}
\label{fppperp}
\setlength{\epsfxsize=0.75\textwidth}
\centerline{\epsffile{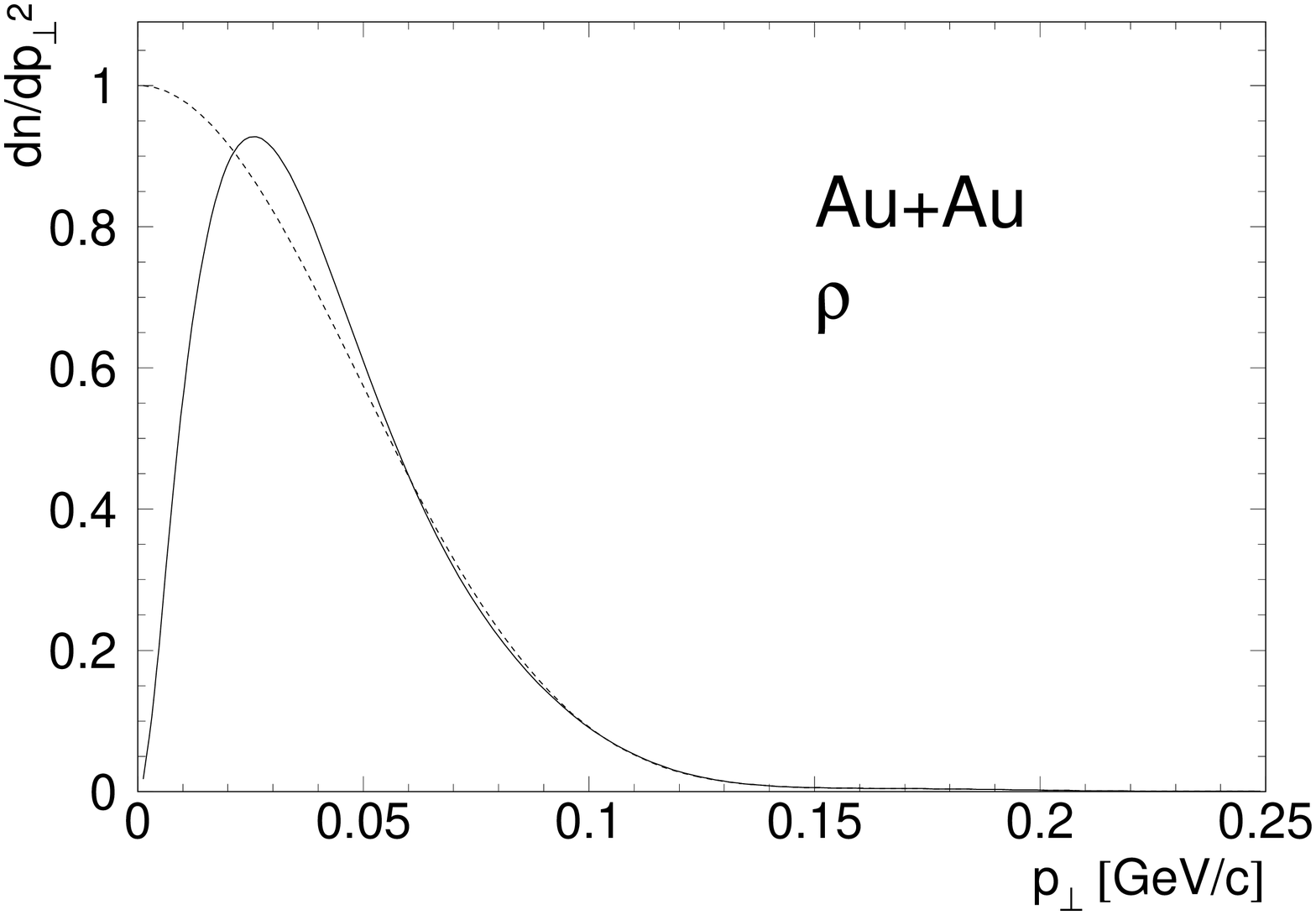}}
\caption[]{Perpendicular momentum spectra for $\rho^0$ production at RHIC, at
$p_z=0$, for gold on gold collisions at a center of mass energy of 200
GeV per nucleon.  Plotted are $dN/dp_T$, with and without
interference.  The curves are normalized to 1 for $p_T=0$ and no
interference.  The calculation assumes that the impact parameter is
not measured, so the interference is washed out, except 
for $p_T < 25$ MeV/c.}
\end{figure}

\end{document}